\documentclass[11pt]{article}
\usepackage{graphicx}
\newcommand{\etal}{{\it et al.}}
\newcommand{\eg}{{\it e.g.}}
\newcommand{\Om}{\Omega}
\newcommand{\om}{\omega}
\newcommand{\si}{\sigma}
\newcommand{\La}{\Lambda}
\newcommand{\De}{\Delta}
\newcommand{\bn}{{\bf n}}
\newcommand{\bx}{{\bf x}}
\newcommand{\be}{\begin{equation}}
\newcommand{\ee}{\end{equation}}
\newcommand{\gsim}{\stackrel{>}{\sim}}
\newcommand{\lsim}{\stackrel{<}{\sim}}
\begin{document}
\title{Physics of  Cosmic Microwave Background Anisotropies and
Primordial Fluctuations}
\author{Ruth Durrer\\
D\'epartement de Physique Th\'eorique, Universit\'e de
Gen\`eve,\\  Quai E. Anserment 24, CH-1211 Gen\`eve, Switzerland} 
\maketitle

%

\begin{abstract}
The physics of the origin and evolution of CMB anisotropies is described, 
followed by a critical
discussion of the present status of cosmic parameter estimation with 
CMB anisotropies.
\end{abstract}


\section{Introduction}
The discovery of anisotropies in the cosmic microwave background (CMB)
by the COBE satellite in 1992~\cite{COBE1,COBE2} has
stimulated an enormous activity in this field which has culminated
recently with the high precision data of the BOOMERanG,
MAXIMA-I and DASI experiments \cite{B98,Max,B2,M2,DASI}. The CMB is 
developing into the most
important observational tool to study the early Universe. Recently,
CMB data has been used mainly to determine cosmological
parameters for a fixed model of initial fluctuations, namely scale
invariant adiabatic perturbations. In my talk I  outline this
procedure and present some results. I will also mention the problem 
of degeneracies and indicate how these are removed by  measurements 
of the CMB polarization or other cosmological data. Finally, I include 
a critical discussion of the model assumptions which enter the parameter
estimations and will show in an example what happens it these
assumptions are relaxed. 

In the next section we discuss in some detail the physics of the CMB. Then we 
investigate how CMB anisotropies depend on cosmological parameters. We also 
discuss degeneracies. In Section~4 we investigate the model dependence of
the 'parameter estimation' procedure. We end with some conclusions.

\section{The physics of the CMB}
Before discussing the possibilities and problems of parameter estimation using
CMB anisotropy data I want to describe the physics of these anisotropies. As
CMB anisotropies are small, they can be treated nearly completely within 
linear cosmological perturbation theory. Effects due to non-linear clustering 
of matter, like {\em e.g.} the Rees-Sciama effect, the Sunyaev-Zel'dovich effect
or lensing are relevant only on very small angular scales ($\ell\gsim 1000$) 
and are not discussed here.

Since the CMB anisotropies are a function on a sphere, they can be expanded
in spherical harmonics,
\be
{\De T\over T_0}(\bn) = \sum_{\ell=1}^\infty\sum_{m=-\ell}^{m=\ell}
	 a_{\ell m}Y_{\ell m}(\bn)~,
\ee
where $\De T=T-T_0$ and $T_0$ is the mean temperature on the sky.
The CMB power spectrum $C_\ell$ is the ensemble average of the coefficients
$a_{\ell m}$,
\[ C_\ell = \langle|a_{\ell m}|^2\rangle ~. \]
If the fluctuations are statistically isotropic, the $C_\ell$'s are 
independent of $m$ and if they are Gaussian all the statistical information 
 is contained in the power spectrum. The relation between 
the power spectrum and the two point correlation function is given by
\be
\left \langle{\De T\over T_0}({\bn_1}){\De T\over T_0}({\bn_2})\right\rangle =
 {1\over 4\pi} \sum_\ell (2\ell+1)C_\ell P_\ell(\bn_1\cdot\bn_2)~.
\ee
In a real experiment, unfortunately, we have only one universe and one sky
at our disposition and can therefore not measure an ensemble average. In 
general, one assumes statistical isotropy and sets
\[  C_\ell \simeq C_\ell^{obs} ={1\over 2\ell+1}\sum_m |a_{\ell m}|^2 ~. \]
In the ideal case of full sky coverage, this yields an average on $2\ell + 1$
numbers (note that $a_{\ell m}=a_{\ell -m}^*$). If the temperature 
fluctuations are Gaussian, the observed mean  deviates from the ensemble 
average  by about 
\be {\sqrt{(C_\ell^{obs}- C_\ell)^2} \over C_\ell} \simeq 
	\sqrt{2\over 2\ell + 1}~.  \label{2cv}
\ee
This fundamental limitation of the precision of a measurement which is 
important especially for low multipoles is called cosmic variance. 
In practice one never has complete sky coverage and the cosmic variance of 
a given experiment is in general substantially larger than the value given 
in Eq.~(\ref{2cv}).

Within linear perturbation theory one can split perturbations into scalar, 
vector and tensor contributions according to their transformation properties 
under rotation. The different components do not mix. Initial vector 
perturbations rapidly decay and are thus usually neglected. Scalar and 
tensor perturbations contribute to CMB anisotropies. After recombination of
electrons and protons into neutral hydrogen, the universe becomes transparent 
for CMB photons and they
move along geodesics of the perturbed Friedman geometry. Integrating
the perturbed geodesic equation, one obtains the following expressions 
for the temperature anisotropies of scalar ($s$) and 
tensor ($t$) perturbations
\begin{eqnarray}
\left({\De T\over T}\right)^{(s)}(\eta_0,\bx_0,\bn) &=& {1\over 4}D_r(\eta_{dec}, \bx_{dec})+ 
 v_i(\eta_{dec}, \bx_{dec})n^i
 +(\Phi-\Psi)(\eta_{dec}, \bx_{dec}) \nonumber \\
 && -\int_{\eta_{dec}}^{\eta_0}
  (\dot\Phi-\dot\Psi)(\eta,\bx(\eta))d\eta~,  \label{ani}\\
\left({\De T\over T}\right)^{(t)}(\eta_0,\bx_0,\bn) &=& 
	-\int_{\eta_{dec}}^{\eta_0}\dot h_{ij}(\eta,\bx(\eta))n^in^jd\eta
	\label{tensor} ~.
\end{eqnarray}
Here $\eta$ denotes conformal time, $\eta_0$ indicates today 
while $\eta_{dec}$ is the time of decoupling ($z_{dec}\sim 1100$)
and $\bx(\eta)$ is the comoving unperturbed photon position 
at time $\eta$, $\bx(\eta)=\bx_0- \bn(\eta-\eta_0)$ for a flat universe,   
$\bx_{dec}=\bx(\eta_{dec})$. The above
 expression for the temperature anisotropy is written in gauge-invariant 
form~\cite{d90}. The variable $D_r$ represents the photon energy
density fluctuations, $v_i$ is the baryon velocity field and $\Phi$ and 
$\Psi$ are the Bardeen potentials, the scalar degrees of freedom for metric 
perturbations of a Friedman universe~\cite{Bardeen}. For perturbations 
coming from ideal fluids or non-relativistic matter $\Psi=-\Phi$ is simply
the Newtonian gravitational potential.

\subsection{The Sachs Wolfe effect}
On large angular scales, the dominant contributions to the power spectrum 
for scalar perturbations come from the first term and the Bardeen potentials.
The integral is often called the 'integrated Sachs Wolfe effect' (ISW) while 
the first and third terms of Eq.~\ref{ani} are the 'ordinary Sachs Wolfe 
effect' (OSW).
In the general case this split is purely formal, but in a matter dominated 
universe with critical density, $\Om_m=1$, the Bardeen potentials are time 
independent and the ISW contribution vanishes.

For adiabatic fluctuations in a matter dominated universe, one has 
${1\over 4}D_r={1\over 3}D_m ={5\over 3}\Psi$. Together with $\Phi=-\Psi$ 
this yields the original formula of Sachs and\\  Wolfe~(1967):
\[ \left({\De T\over T}\right)^{SW} = -{1\over 3}\Psi~. \]

Tensor perturbations (gravity waves) only contribute on large scales, 
where metric perturbations are most relevant. Note the similarity of the 
tensor contribution to the ISW term which has the same origin.

\subsection{Acoustic oscillations and the Doppler term}
Prior to recombination, photons, electrons and baryons form a tightly 
coupled fluid.
On sub-horizon scales this fluid performs acoustic oscillations driven by 
the gravitational potential. The wave equation in Fourier space is 
\begin{eqnarray}
\dot{D} +3(c_s^2 -w){\dot a\over a}D +(1+w)k^2V &=&0 \label{ac1}\\
\dot V +{\dot a\over a}(1-3c_s^2)V -{c_s^2\over w+1}D &=& 
	\Psi-3c_s^2\Phi \label{ac2}
\end{eqnarray}
where $w=p/\rho$ and $c_s^2={\dot p/ \dot \rho}$ is the adiabatic sound speed.
Since before recombination, the baryon photon fluid is dominated by radiation 
we have $w\simeq c_s^2\simeq 1/3$.The system~(\ref{ac1},\ref{ac2}),
which is a pure consequence of energy momentum conservation for
the baryon photon fluid, can be combined to a second order wave equation for 
$D$. On very large, super-horizon scales, $k\eta\ll 1$ the oscillatory term 
can be neglected and $D$ remains constant. Once $k\eta\gsim 1$ $D$ begins to 
oscillate like an acoustic wave. For pure radiation, $c_s^2=w=1/3$ the 
damping term vanishes and the amplitude of the oscillations remains constant.
At late times there is a slight damping of the oscillations. 

If adiabatic perturbations have been created during an early inflationary 
epoch, the waves are in a maximum as long as $k\eta\ll 1$ and perturbations 
with a given wavenumber all start oscillating in phase. At the moment of 
recombination, when the photons become free and the acoustic oscillations 
stop, the perturbations of a given wave length thus have all the same 
phase. As each given wave length
is projected to a fixed angular scale on the sky, this leads to a 
characteristic series of peaks and troughs in the CMB power spectrum. The 
first two terms in Eq.~(\ref{ani}) are responsible for these acoustic peaks.

\begin{figure}[ht]
\vspace*{-0.4cm}
\centerline{\includegraphics[clip=,width=0.5\textwidth]{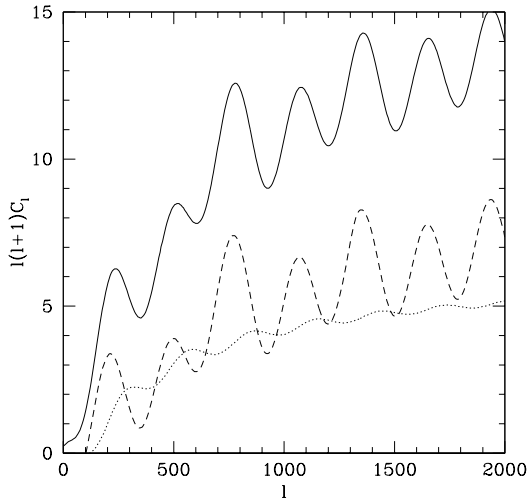}}
\caption{The contribution from the photon density fluctuations 
($D_r$-term, dashed line), from the Doppler term 
(${\bf v}\cdot \bn$, dotted line) and their sum (solid line) are shown. 
The peaks clearly follow the acoustic contribution while the Doppler 
term dominates in the first few minima. Silk damping is not included here.}
\label{fig:acc}
\end{figure}

In Fig.~\ref{fig:acc} we show the density and the velocity terms as well as 
their sum. The density term is often called the 'acoustic term' while the 
velocity term is the 'Doppler term'. It is clearly wrong to call the peaks 
in the CMB anisotropy spectrum 'Doppler peaks' as the Doppler term actually 
is close to a minimum at the position of the peaks! We therefore call 
them acoustic peaks.

\subsection{Silk damping}
So far we have neglected that the process of recombination takes
a finite amount of time and the 'surface of last scattering' has a finite 
thickness. In reality the transition from perfect fluid coupling with a very 
short mean free path to free photons with mean free path larger than the size 
of the horizon takes a certain time during which photons can diffuse out of 
over--densities into under--densities. This diffusion damping or Silk 
damping~\cite{Silk} exponentially reduces CMB anisotropies on small 
scales corresponding to $\ell\gsim 800$. The precise damping scale depends
on the amount of baryons in the universe.

In addition to Silk damping, the finite thickness of the recombination shell
implies that not all the photons in the CMB have been emitted at exactly 
the same moment and therefore we do not see all the fluctuations precisely 
in phase. This 'smearing out' also leads to damping of CMB anisotropies on 
about the same angular scale as Silk damping.

To calculate these phenomena with good precision one has to compute
the process of recombination numerically and integrate the photon Boltzmann
equation. Since a couple of years there are public codes 
available~\cite{CMBfast,CAMcode} which compute the CMB anisotropies 
numerically with a precision of about $1$\%. 

\subsection{Polarization}
There is an additional phenomenon which we have not considered so far: 
Non-relativistic Thompson scattering, which is the dominant scattering 
process on the surface of last scattering, is anisotropic. The scattering
cross section for photons  polarized in the scattering plane is \cite{J}
\[ \si_{//} = {3\si_T\over 8\pi}\cos^2\theta ~,\]
while the cross section for photons polarized normal to the plane is
\[ \si_{\perp} = {3\si_T\over 8\pi} ~.\]
Here $\si_T$ is the Thomson cross section and $\theta$ is the scattering angle.
Therefore, even if the incoming radiation is completely unpolarized, if its
intensity
is not perfectly isotropic (actually if it has a non-vanishing quadrupole) the 
outgoing radiation will be linearly polarized. There exist two types of 
polarization signals: the so called $E$-type polarization which has positive 
parity, and $B$-type polarization which is parity odd. Scalar perturbations 
only produce $E$-type polarization, while tensor perturbations, gravity 
waves, produce both, $E$- and $B$-type. Thomson scattering
never induces circular polarization.

\begin{figure}[ht]
\centerline{\includegraphics[clip=,width=0.6\textwidth]{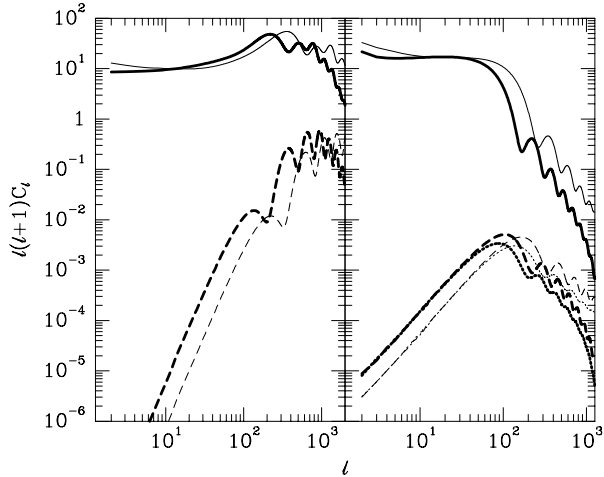}}
\caption{The scalar (left) and tensor (right) CMB anisotropy and 
polarization spectra. Solid lines show  the temperature anisotropy,
dashed lines  $E$-polarization and  dotted lines  $B$-polarization. 
 The thick lines represent a model with critical density, $\Om_0=1$
while the thin lines come from an open model, $\Om_0=0.4$. The normalization
is arbitrary. Figure  from Hu \etal~(1998).}
\label{fig:CMBpol}
\end{figure}

A more detailed treatment of polarization of CMB anisotropies can be found 
\eg ~in~\cite{SWH}. A typical CMB anisotropy and polarization spectrum as it
is expected from inflationary models is shown in Fig.~\ref{fig:CMBpol}. 
Polarization of the CMB has not yet been observed. The best existing limits 
are on the order of a few$\times 10^{-6}$. There is hope that the next 
Boomerang flight (planned for December 2001) or the MAP satellite~\cite{map},
which has been launched successfully in June 2001, will detect polarization.

\section{Cosmological parameters and degeneracy}
\begin{figure}[ht]
\centerline{\includegraphics[clip=,width=0.52\textwidth]{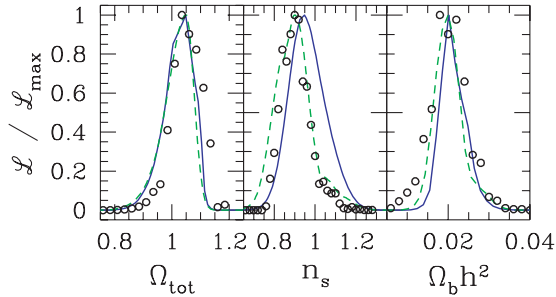}}
\caption{The likelihood curves derived from the BOOMERANG and COBE/DMR 
data sets for the variables $\Om_{tot}=\Om_0$,  $n_s$ and  
$\Om_bh^2=\om_b$ for a model with
purely scalar adiabatic fluctuations are shown. The solid lines are 
marginalized over the other variables while for the dashed lines (and the 
open circles) the maximum likelihood point in the other variables is chosen;
from de Bernardis \etal~(2001).}
\label{fig:param}
\end{figure}

In the simplest models for structure formation where adiabatic Gaussian
perturbations 
are created during an inflationary phase, initial perturbations are 
characterized by two to four numbers: The amplitudes and spectral indices 
of scalar and tensor perturbations. Apart from these data characterizing the 
initial conditions, the resulting CMB anisotropies depend only on the 
cosmological parameters of the underlying model, the matter density 
parameter, $\Om_m$, the cosmological constant, $\Om_\La$,  curvature, 
$\Om_K =1-\Om_0$, the Hubble parameter, $h = H_0/(100{\rm km/s/Mpc})$, the 
(reduced) baryon density $\om_b=\Om_bh^2$, the reionization history, which is 
usually cast into an effective depth to the last scattering surface, 
$\tau_c$, and a few others. Therefore, if the model of 
structure formation is a simple adiabatic inflationary model, CMB anisotropies
can be used to determine cosmological parameters. The presently available
data have been used for this goal in numerous papers and slightly different 
approaches have led to slightly different but, within the still considerable
 error bars, consistent results (see e.g.~\cite{B98,Pyke,Lange} and 
many others). As an example we show the results of~de Bernardis \etal~(2001).
In Fig.~\ref{fig:param} the likelihood functions for the total 
density parameter,
$\Om_0$, the scalar spectral index, $n_s$, and the baryon density, $\om_b$,
as obtained from the COBE DMR and the BOOMERANG data are shown \cite{like}. 
An adiabatic model with purely scalar perturbations, with $0.45<h<0.95$
and with an age larger than 10Gyr has been assumed for the determination of 
the likelihoods. The solid lines, which have been obtained by marginalization
over all the parameters not shown on the panel, are the most relevant. 
They imply $\Om_0 =1.02\pm 0.06$, $n_s=1.02\pm 0.1$ and 
$\om_b=0.024\pm0.005$. The latter value coincides most remarkably with the 
completely independent determination from nucleosynthesis result~\cite{nuc} 
which yields $\om_b=0.019\pm0.02$.

The most interesting outcome from these parameter estimations 
is that if initial perturbations are adiabatic, the Universe is very close 
to flat. Together with the cluster data which indicate $0.1\le \Om_m\le 0.3$ 
this suggests, completely independent from the supernova results, that the
density of the universe is dominated by a non-clustered form of dark energy,
\eg~a cosmological constant with $\Om_\La \sim 0.7$.

\begin{figure}[ht]
\centerline{\includegraphics[clip=,width=0.49\textwidth]{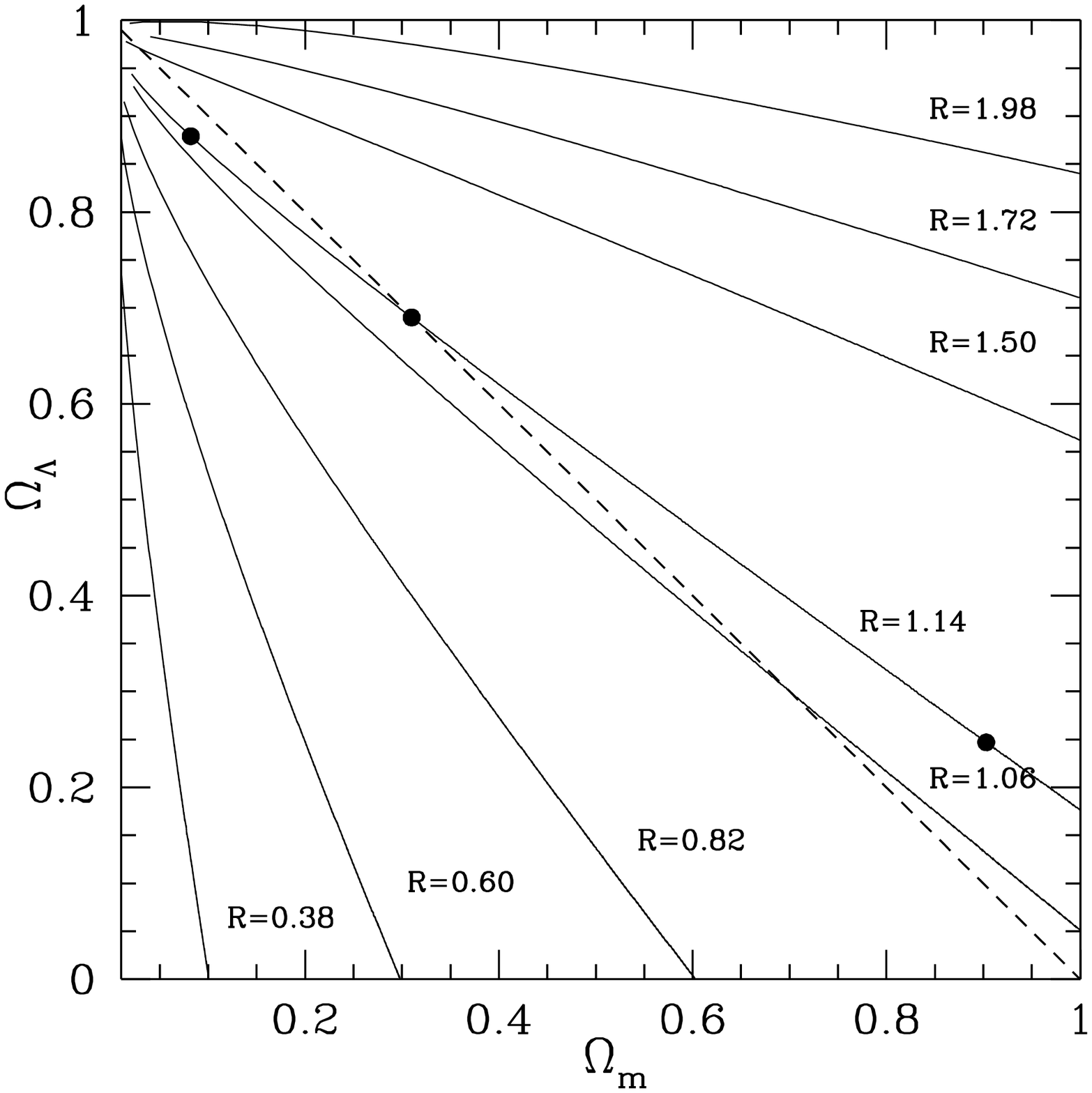}\includegraphics[clip=,width=0.49\textwidth]{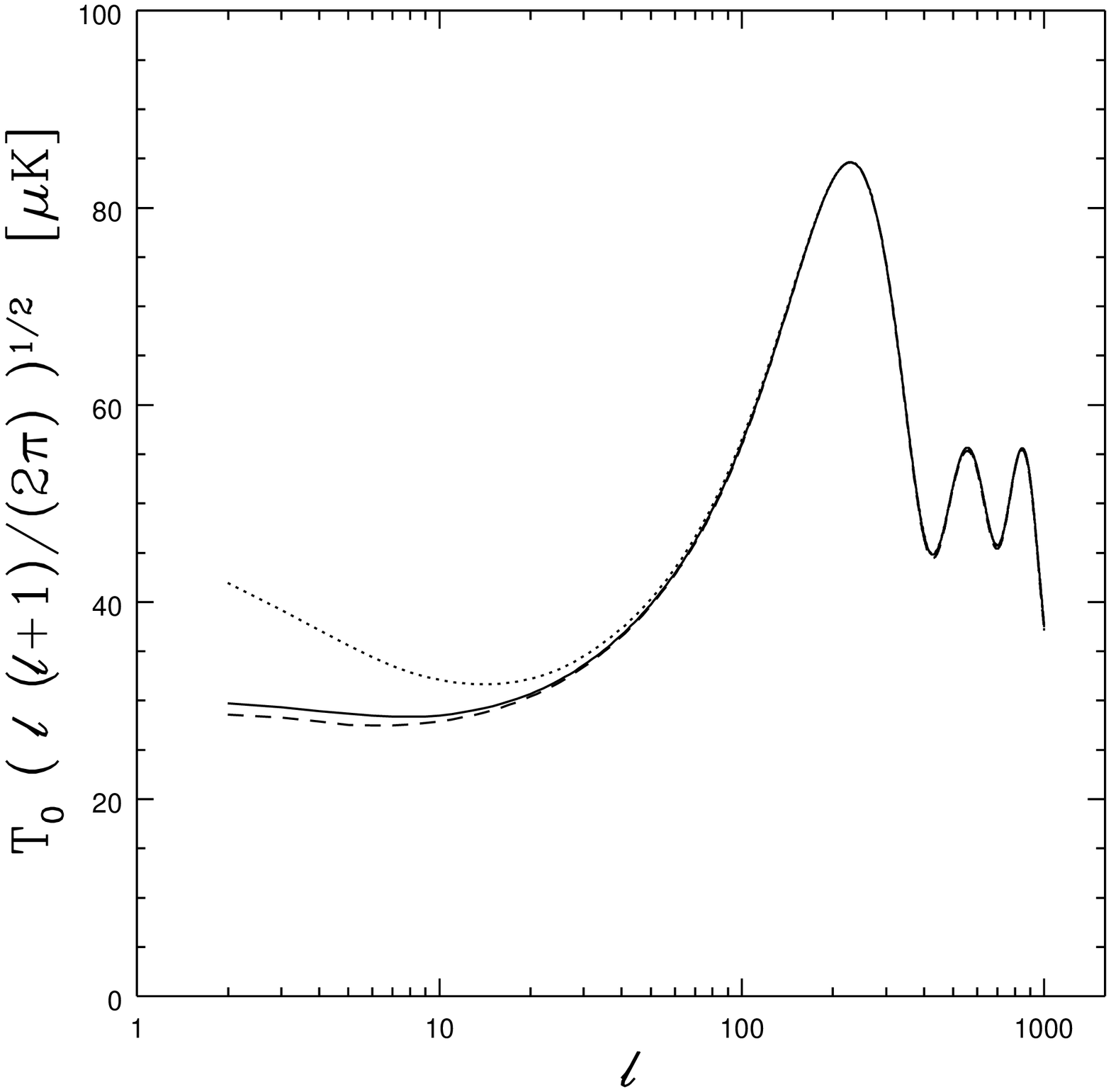}}
\caption{{\bf left:} The lines of constant $R$ are shown in the 
$\Om_\La$--$\Om_m$
plane. The values $\Om_\La,\Om_m$ for which the CMB anisotropy spectra are
shown right are indicated as black dots.
\hspace{3cm}
{\bf right:}Three CMB anisotropy spectra with different values of 
$\Om_\La$ and $\Om_m$ but fixed $R$ are shown. For $\ell\gsim 50$ these 
spectra are clearly degenerate.The solid line represents a flat model, while 
the dotted line corresponds to a closed and the dashed line to an open 
universe,  from Trotta~(2001).}
\label{fig:degeneracy}
\label{fig:olom}
\end{figure}

However promising this  procedure is, it is important to keep in mind that 
there are certain exact degeneracies in the CMB data which cannot be removed
by CMB data alone. Let us consider, for example, the parameters $\Om_m,
\Om_\La,\Om_b, h$. Apart from the ISW contribution which is relevant only 
at low values $\ell$ where cosmic variance prohibits a precise determination,
the CMB anisotropies depend on these parameters only via the baryon density,
$\om_b$, the matter density $\om_m=\Om_mh^2$ and the angular diameter distance
given by $d_{dec}=\chi(\eta_0-\eta_{dec})$, where 
\[ \chi(y) =\left\{\begin{array}{ll}
   \sin(y) & \mbox{ if }~~ K>0 \\
   \sinh(y) & \mbox{ if }~~ K<0 \\
   y& \mbox{ if }~~ K =0, \end{array}\right.
\]
and
\[y=\eta_0-\eta_{\rm dec} =\sqrt{|\Omega_K|}\int_0^{z_{dec}}
	{dz\over [\Omega_m(1+z)^3 + \Omega_K(1+z)^2
	+\Omega_\Lambda]^{1/2}}.
\]
The CMB anisotropies for $\ell\gsim 50$ only depend on the following three 
combinations of the four parameters considered:
$ R \equiv d_{dec}(\Om_\La=0,\Om_m=1)/ d_{dec}(\Om_\La=0,\Om_m),~ \om_b$
and $\om_m$.
In Fig.~\ref{fig:olom} lines of constant $R$ are indicated in the 
$\Om_\La$--$\Om_m$ plane.

The degeneracy is shown in Fig.~\ref{fig:degeneracy} for a fixed value of $R$
but different points in the $\Om_\La$--$\Om_m$ plane.

It is hence not possible to determine all four parameters  $\Om_m,
\Om_\La,\Om_b, h$ with good accuracy from CMB data alone. There exist also 
other degeneracies, \eg~between the spectral index and the epoch of 
reionization or the amplitude of tensor perturbations.

We therefore consider it very important that CMB anisotropy measurements are 
complemented with other, more direct methods to measure cosmological 
parameters so that this degeneracies are broken, and also to obtain a 
comfortable degree of redundancy.

\section{Model dependence}
\begin{figure}[ht]
\centerline{\includegraphics[clip=,width=0.6\textwidth]{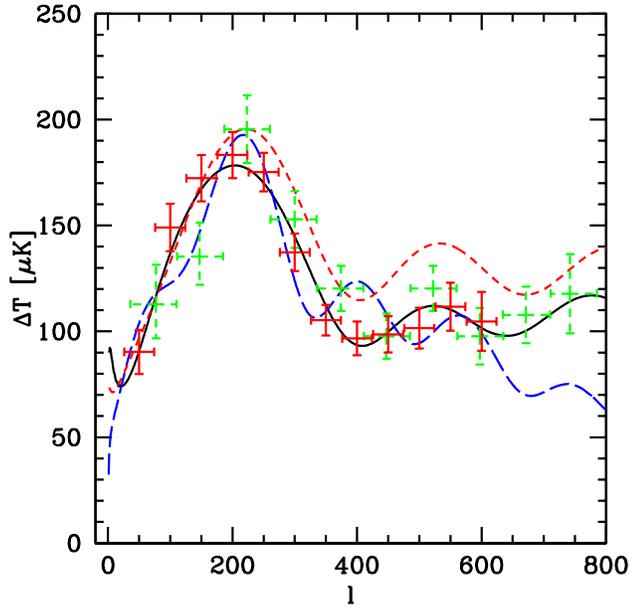}}
\caption{CMB anisotropies for three different models are shown, together 
with the Boomerang (solid, red) and Maxima (dashed, green) data. The short 
dashes show a standard adiabatic inflationary model, the long dashes show a
closed isocurvature model, while the solid line shows another so called
 'scaling seed' model. Figure from Durrer \etal~(2001b).}
\label{fig:models}
\end{figure}

Apart from the degeneracies mentioned in the previous section, the 
cosmological parameters inferred form CMB anisotropies very strongly depend 
on the model assumptions. For example, in the case of isocurvature instead of 
adiabatic initial perturbations, for a model with critical density the first
acoustic peak is at $\ell\sim 350$, and a peak at $\ell\sim 210$ indicates a
closed universe. However, a closed model with isocurvature perturbations
has acoustic peaks which are narrower and more closely spaced than those 
seen in the data. One such model, together with the data, is shown in 
Fig.~\ref{fig:models}. More details can be found in~\cite{DKM1,DKM2}.

\begin{figure}[ht]
{\centerline{\hspace*{0.5cm} \includegraphics[clip=,angle=-90,width=0.57\textwidth]{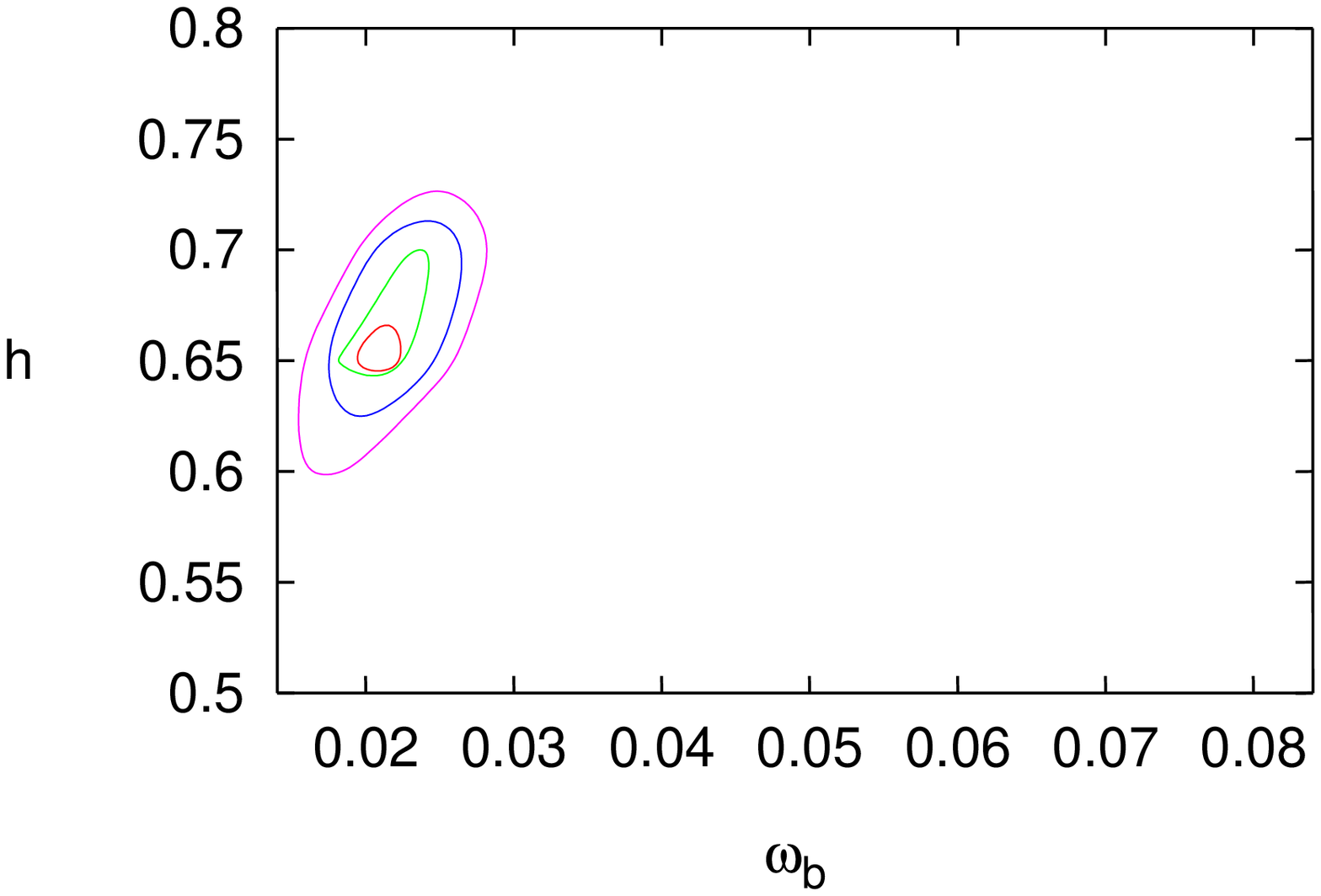} \includegraphics[clip=,angle=-90,width=0.57\textwidth]{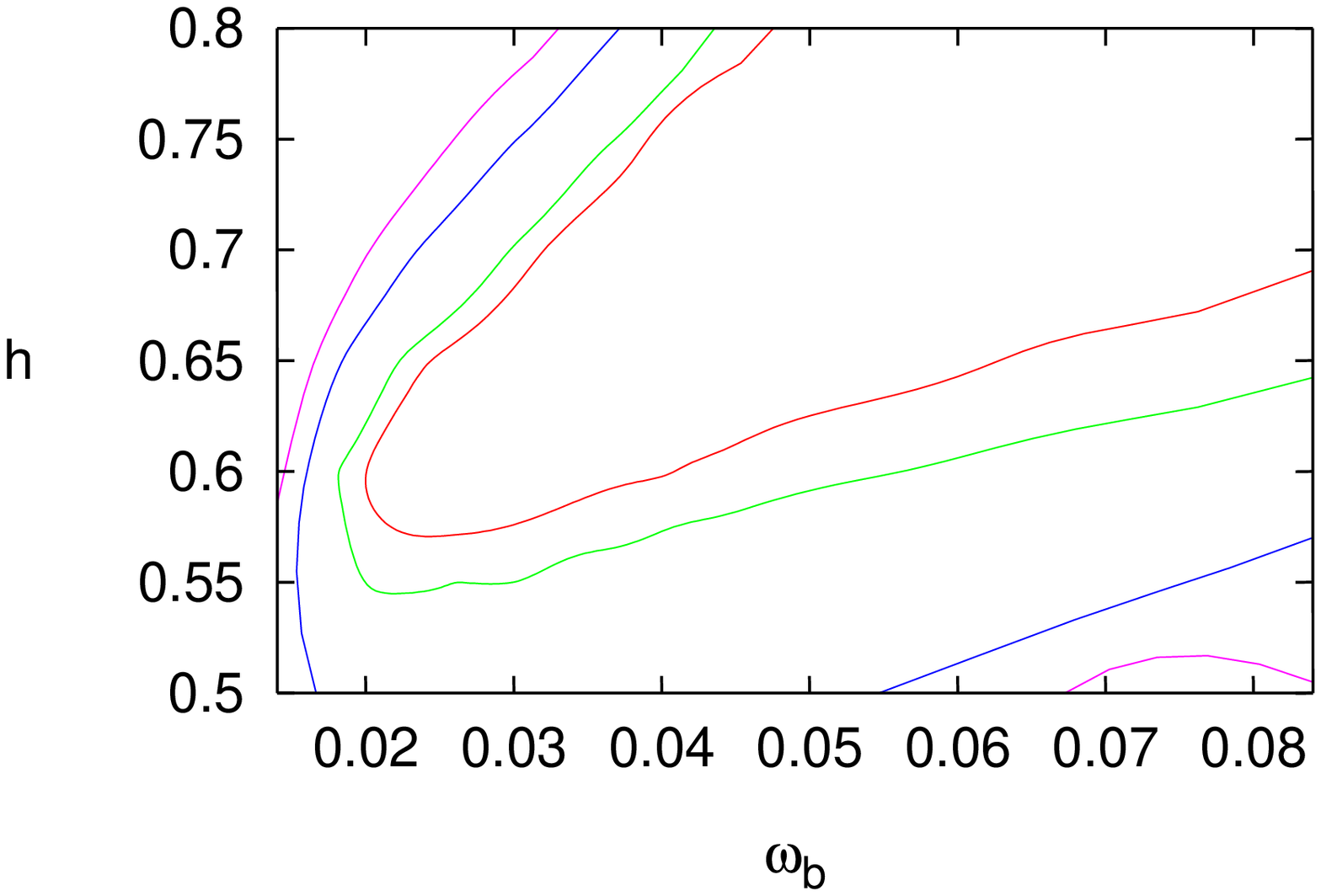}}}
\caption{ The $h,\om_b$ confidence contours as obtained from the Boomerang 
data~\protect\cite{B2}, when allowing only for the adiabatic mode 
(left) and when allowing for a general correlated mixture of the adiabatic 
and isocurvature modes (right). Figure from  Trotta \etal~(2001). }
\label{fig:iso}
\end{figure}

Very generic initial conditions for a universe with dark matter, photons,
baryons and neutrinos are  combinations of the adiabatic mode and four 
different isocurvature modes which may or may not be correlated~\cite{Turok}. 
The initial conditions 
are then specified by a $5\times5$ positive definite matrix, the correlation 
matrix of the different modes. It is interesting to compare parameter 
estimation when allowing for this more generic initial conditions to the
parameters obtained from the data when allowing only for the adiabatic mode.
As an example we show the confidence ranges in the $h,\om_b$ plane for
both cases in Fig.~\ref{fig:iso} (see Trotta \etal~(2001)).

Clearly, once allowing for isocurvature modes one cannot obtain anymore 
reasonable upper limits for $\om_b$ or $h$. What I find even more 
interesting here
is that once we require $\om_b\sim 0.02\pm 0.002$ due to the nucleosynthesis 
constraint and $h\sim 0.65\pm 0.1$ as favored by several independent 
estimates, we find that the isocurvature content in the initial conditions 
has to be relatively modest, $\lsim 30$\%.

Nevertheless, I believe that the above makes it clear that estimation of 
cosmic parameters by CMB anisotropies is strongly model dependent.

\section{Conclusions}
In this talk I have discussed the physics of CMB anisotropies and what we 
can learn from them. I have been relatively critical in my account of 
cosmic parameter estimation from CMB anisotropies. This because in the very 
abundant literature on the subject, little emphasis is given on the model
dependence of this way of  'measuring' cosmological parameters. Clearly every
measurement in physics and even more so in cosmology depends on the 
underlying theory. But usually the theory has been tested before in many 
different setups, while in cosmology, CMB anisotropies are probably the best
experimental data to test theories for cosmological initial perturbations,
{\em i.e.} to investigate cosmological perturbations at a very early stage. 
Therefore I find it to some extent a waste if one uses these data simply to 
determine a few numbers which can also be measured much more directly 
(\eg~by kinematic measurements with SNeIa's).

On the other hand, it is intriguing how well the present CMB data can be 
fit by the simplest adiabatic model of scalar perturbations with 
cosmological parameters well within the range obtained by other measurements.
\vspace{1cm}

I thank the organizer of the workshop for providing such an
interesting conference and a stimulating atmosphere for discussions. I have
profited from many discussions with colleagues, especially Paolo de Bernardis,
Pedro Ferreira, Martin Kunz, Alessandro Melchiorri, Alain Riazuelo, Roberto 
Trotta and Neil Turok. I'm grateful to Norbert Straumann for carefully 
reading the first version of this contribution. I also thank the Institute 
for Advanced Study, where this paper was completed, for hospitality.

\end{document}